\documentclass[conference,letterpaper]{IEEEtran}
\usepackage{fancyhdr}
\usepackage{latexsym}
\usepackage{epsfig}
\usepackage{psboxit}
\usepackage{wrapfig}
\usepackage{subfigure}
\usepackage{algorithm}
\usepackage{algorithmic}
\usepackage{amsfonts}
\usepackage{graphicx}

\usepackage{multirow}

\usepackage{fancyhdr}

\begin{document}
\title{An Aggregation-Based Overall Quality Measurement for Visualization}

\author{\IEEEauthorblockN{Weidong Huang}
\IEEEauthorblockA{CSIRO, Australia\\
Tony.Huang@csiro.au}
%\and
%\IEEEauthorblockN{Authors Name/s per 2nd Affiliation}
%\IEEEauthorblockA{line 1 (of Affiliation): dept. name of organization\\
%line 2: name of organization, acronyms acceptable\\
%line 3: City, Country\\
%line 4: e-mail address if desired}
}

\maketitle
\thispagestyle{plain}

% insert page header and footer here for IEEE PDF Compliant
\fancypagestyle{plain}{
\fancyhf{}	% clear all header and footer fields
\fancyfoot[L]{}
\fancyfoot[C]{}
\fancyfoot[R]{}
\renewcommand{\headrulewidth}{0pt}
\renewcommand{\footrulewidth}{0pt}
}

\pagestyle{fancy}{
\fancyhf{}
\fancyfoot[R]{}}
\renewcommand{\headrulewidth}{0pt}
\renewcommand{\footrulewidth}{0pt}

\begin{abstract}
Aesthetics are often used to evaluate the quality of graph drawings. However, the existing aesthetic criteria are useful in judging the extents to which a drawing conforms to particular drawing rules. They have limitations in evaluating overall quality. Currently the overall quality of graph drawings is mainly evaluated based on personal judgments and user studies. Personal judgments are not reliable, while user studies can be costly to run. Therefore, there is a need for a direct measure of overall quality. This measure can be used by visualization designers to quickly compare the quality of drawings at hand at the design stage and make decisions accordingly. In an attempt to meet this need, we propose a measure that measures overall quality based on aggregation of individual aesthetic criteria. We present a user study that validates this measure and demonstrates its capacity in predicting the performance of human graph comprehension. The implications of the proposed measure for future research are discussed.
\end{abstract}


% for peerreview papers, inserts a page break and creates the second title.
% Will be ignored for other modes.
\IEEEpeerreviewmaketitle

\section{Introduction} \label{sec:intro}

In drawing graphs, one of the issues we face is how to lay them out as empirical research has shown that layout affects how a graph is perceived (e.g.,~\cite{huang05}). A range of rules for laying out graphs, or aesthetics, have been proposed in the field of graph drawing. Examples of such aesthetics include minimum crossings and maximum symmetries. It is commonly believed that drawings satisfying these aesthetics are more effective. As a result, aesthetics have been widely used as quality criteria in evaluating automatic graph drawing algorithms and interfaces of graph visualization systems (e.g.,~\cite{dunne,didimo}).

However, the existing aesthetic criteria are useful in judging the extents to which a drawing conforms to specific drawing rules. They have limitations in evaluating the overall quality. Part of the reason is that aesthetics often conflict with each other and cannot be implemented in full at the same time. This conflicting relationship affects current practices of graph drawing greatly. On the one hand, many algorithms for automatic graph drawing are designed to optimize only one or two aesthetics, and different algorithms focus on different aesthetics~\cite{di}. This makes it difficult for an algorithm user to choose which algorithm to use when he or she has more than one candidate algorithm at hand. We do not know whether an algorithm that is to minimize the number of crossings will produce better drawings than another algorithm that is to maximize symmetries. On the other hand, it is generally acknowledged that the best layout is the balance of aesthetics. However, seeking a compromise between a set of aesthetics only gives us a better chance of producing good drawings~\cite{huang10}. We still do not know whether a drawing produced based on one set of aesthetics is better than another drawing that is produced based on a different set of aesthetics. 

Due to the lack of appropriate measures, the overall quality of graph visualizations is often evaluated based on personal judgments and user studies. However, personal judgments are subjective and are not reliable, while user studies can be costly to run. Therefore, there is a need for a reliable and objective measure so that we can evaluate overall quality at the early design stage of a visualization process. This measure will help visualization designers to quickly judge or compare the quality of the drawings in consideration and make decisions accordingly. 

In an attempt to meet this need, we propose an overall quality measure of layout. This measure takes into account individual aesthetic criteria and gives a single numerical value. In this paper, we explain how this measure is computed. We present a study to validate the quality measure and to demonstrate its capability in predicting the performance of human graph comprehension.

\section{Related work} \label{sec:related}

Two main approaches have been used in evaluating graph drawings: 1) approaches based on computational measures such as commonly used aesthetics and specifically developed measures; and 2) approaches based on empirical measures such as expert opinion, user preference and task performance. In this section, selected studies are given as examples. 

Didimo et al.~\cite{didimo} conducted a study that evaluated different graph drawing algorithms. The quality of the resultant drawings was compared based on the extent to which they conformed to each of a set of aesthetics. The aesthetics used for the comparison purpose included number of crossings, crossing angle resolution, geodesic edge tendency and vertex angles. The results indicated that some algorithms had a better trade-off between these criteria than other algorithms. 

Brandes and Pich~\cite{brandes08} compared distance-based graph drawing algorithms. In this study, the drawing quality was measured as how well the Euclidean distance between any two nodes represented their graph-theoretic distance. The results suggested that minimization of weighted stress yielded better layout than force-directed placement in terms of pairwise distances between nodes. Similar quality measurements were also developed in other studies to evaluate whether drawings have specific layout features (e.g.,~\cite{zai}). 

While these computational approaches evaluate quality of specific visual properties of drawings directly, empirical methods have been used to evaluate overall quality. Huang et al.~\cite{huang05} conducted a study that evaluated quality of sociograms that were drawn based on different drawing conventions and edge crossings. Subjects were asked to perform  network-related tasks and indicate their preference for drawings conventions and crossings. Task performance including response times and error rates were also recorded. One of their findings was that a drawing that was the most preferred might not be the one that induced the best performance. Other studies in this category include ~\cite{arch12a,korner02,dwyer09}. Evaluating quality based on expert opinion, Hachul and Junger~\cite{hac} conducted a study that compared algorithms for drawing general large graphs. Rather than using a quantitative measurement, the authors evaluated the drawing quality by commenting how well the layout displayed the graph structure. 

The above-mentioned approaches are useful at certain circumstances of a visualization process and for specific purposes. However, a direct, empirically validated measure is more preferable for a quick assessment of overall quality.

\section{The Proposed Quality Measure}

Aesthetics used in graph drawing have either been empirically validated or widely acknowledged for their association with human graph comprehension~\cite{purchase95,di}. Further, when it comes to the performance of human graph comprehension, each aesthetic has a role to play, and it is the joint effect of these aesthetics that is more relevant. Based on the overall quality measure proposed by Huang et al.~\cite{huang12}, we measure overall visual quality ($y$) as an aggregation of aesthetics ($x$) as below:

\begin{equation}
\label{eq1}
y = \sum\limits_{i=0}^{n} x_{i}
\end{equation}

For a measure to be useful, it should be objective (give the same value when used by different assessors), reliable (give the same value when used at different times), sensitive to changes (be able to tell the difference when there is a change to the visualization), easy to measure (take only a few steps to compute), be comparable (be able to give continuous numerical values, rather than categorical), and be predictive of task performance.

In addition, although being subjective, it is our belief that for this measure to be generically useful, aesthetics to be considered in equitation~\ref{eq1} should be context or application independent, be applicable to general graphs and reflect specific local features of layout.

Keeping these requirements in mind, we chose the following four most discussed aesthetics as the aesthetic elements of equation~\ref{eq1}:

\begin{enumerate}
\item{Minimize the number of edge crossings ($cross\#$)}
\item{Maximize crossing angle resolution ($crossRes$)}
\item{Maximize node angular resolution($angularRes$)}
\item{Uniformize edge lengths ($uniEdge$)}

\end{enumerate}

Among these aesthetics, $cross\#$ is measured as the number of crossings in the drawing; a smaller value is better. $CrossRes$ is measured as the minimum size of all crossing angles; a larger value is better. $AngularRes$ is measured as the minimum size of angles formed by any two neighboring edges; a larger value is better. $UniEdge$ is measured as the standard deviation of all edge lengths; a smaller value is better. The scores of these aesthetics are measured on different scales. To be able to combine them together, they need to be transformed to $z$ scores first. To be more specific, suppose that we have three drawings of the same graph that we would like to compare, and they have 2, 7, and 6 crossings, respectively. That is, the scores ($x$) of $cross\#$ are 2, 7 and 6. The mean of these three scores ($Mean$) is 5 and their standard deviation ($StDev$) is 2.65. Then, the $z$ score of $cross\#$ for each of these three drawings can be computed as below:

\begin{equation}
\label{eq2}
z_{cross\#}=\frac{x-Mean}{StDev}
\end{equation}

\noindent and it is $-1.14$, $0.76$ and $0.38$, respectively. The other aesthetics can be standardized into $z$ scores in the same way.

In aggregating $z$ scores, the scales must be made going the same direction. That is, higher values are always better. As such, equation~\ref{eq1} can be refined and the overall quality score ($O$) can be computed as below:

\begin{equation}
\label{eq3}
O=-z_{cross\#}+z_{crossRes}+z_{angularRes}-z_{uniEdge}
\end{equation}

\section{Experiment} \label{sec:experiment}

In this section, we describe an experiment that was designed to validate the proposed measure and test its predictability.

\subsection{Design}
To validate the measure, we generated a set of random graphs with similar structures. For each graph, a number of drawings were generated using a specific algorithm so that the relative quality levels between these drawings were known beforehand. The overall quality scores of these drawings were computed using the proposed measure. These computed scores were then tested to see whether they were consistent with the pre-known quality. 

To test the capacity of the proposed measure in predicting task performance, we used a random graph with reasonable size and complexity. This graph was randomly drawn a number of times. However, this time we did not know their overall quality beforehand. Instead, we recruited users to perform a typical graph reading task. We measured task response time, accuracy, cognitive load and \emph{visualization efficiency}~\cite{huang09}. We also measured overall quality of each drawing using the proposed measure. We then ran regression tests to see whether the task performance data were significantly associated with the measured overall quality.

In addition, as mentioned in section 2, task performance measures are often used for quality evaluation. We would like to compare how performance measures and the proposed measure performed in differentiating drawings based on overall quality. We therefore included the drawings for the validity test in our user study. As a result, the experiment included two blocks of the drawings: one block for validity and the other for predictability. The experiment employed a within-subject design.
 
\subsection{Stimuli}

\begin{figure*}[t]
\centering
\includegraphics[width=4.6in]{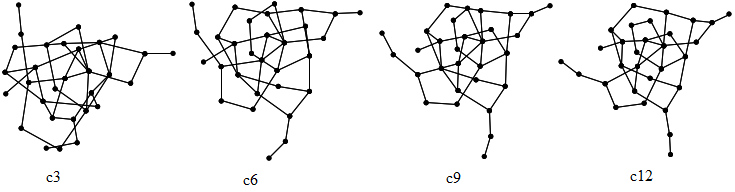}
\caption{An example of the four drawing conditions of a graph for validity} \label{fig:stimuli1}
\end{figure*}

For validity, we generated 20 different graphs based on the Erdos-Renyi model of random graphs [19], with each having 30 nodes and 40 edges. These graphs were then drawn using a force-directed algorithm. The algorithm applies forces on the nodes and edges of a random initial layout, and moves them accordingly. This process is repeated until an equilibrium state is reached. It is known that each time the process is repeated, the overall layout is generally improved. To create experimental conditions, we recorded the layout when the process had been repeated 3000, 6000, 9000 and 12000 times for each graph. As a result, four conditions were obtained: c3, c6, c9 and c12. Each graph had one drawing in each condition and each condition had 20 drawings. The drawing quality improved across the conditions from c3, c6, c9 to c12. Figure 1 shows the four drawings of a graph.

%\begin{figure*}[bp]
\begin{figure*}[t]
\centering
\includegraphics[width=4.6in]{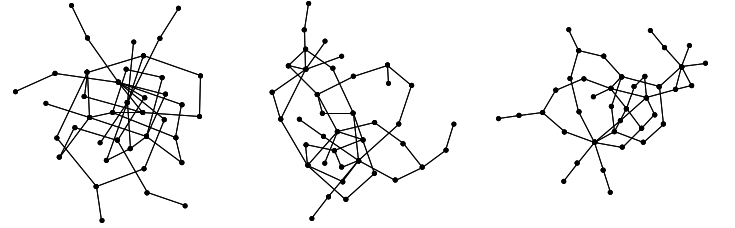}
\caption{Three examples of the thirty drawings of a graph for predictability} \label{fig:stimuli2}
\end{figure*}

For predictability, we used a graph that had 39 nodes and 48 edges. This graph was randomly drawn thirty times using a force-directed algorithm, resulting in 30 drawings in total. Each of these drawings was obtained by using a random combination of a different initial layout, a different number of iterations for convergence and a different set of parameters that were used to define the forces. Figure~\ref{fig:stimuli2} shows three examples of them.

To avoid possible fatigue or boredom caused by too many drawings, the twenty graphs for validity were divided into two halves. Only one half was chosen in an alternating order and the corresponding drawings of the chosen half were used in each trial. As a result, each subject viewed 10 $\times$ 4 (conditions) = 40 drawings for validity and 30 drawings for predictability, that is, 70 stimuli in total.

\subsection{Subjects}

Thirty-five subjects volunteered to participate in the study. These subjects were undergraduate students. All of them had normal or corrected-to-normal vision.

\subsection{The task}

The task was to find the shortest path between two pre-specified nodes. Subjects were required to response by indicating the length of the path. For each graph, the two nodes were randomly chosen with two pre-conditions. The first condition was that there was only one shortest path between them. The second was that the shortest path length was between 3 to 5 inclusive for validity and 4 for predictability. Given a graph, the same path was used across the drawings of it for a subject, while the paths could be different between subjects. Using different paths for different subjects was to ensure that the impact of overall layout, rather than a specific part of the drawing, was reflected in performance data.

\subsection{The online system}

A custom-built system was used to display the drawing images. The system was designed to display the $70$ stimuli in a random order with one constraint and to highlight the two pre-specified nodes as red. The constraint for the randomization of the order was that no two drawings from the predictability block were displayed in a row. It displayed one of the red nodes first. The subject looked at the node and hit a key on the keyboard to have the whole drawing displayed. Then the subject started looking for the answer. Once the answer was found, the subject was required to hit a key immediately. Once the key was hit, the time spent for the answer was recorded and an answer screen was shown. The answer screen had six boxes on the top, with a number near each box representing a possible answer to the task. There were also nine smaller boxes with numbers indicating possible levels of mental effort devoted to the task (from 1 being the lowest to 9 being the highest). The subject was required to respond by clicking on one box of each set. After two answers had been given, the subject hit a key to have the responses recorded and to continue with the next drawing. 

\subsection{Procedure}

Before the experiment, subjects were given time to read the information and tutorial documents, understand the graph concepts and the task, and practice with the system. They were also free to ask questions. Subjects were instructed to perform the task as quickly as possible without compromising accuracy.

Once ready, subjects indicated to the experimenter and the experiment started. At any time when the answer screen was displayed, subjects could have a break as they wished before a key was hit. Therefore, the pace of the experiment was controlled by subjects in order to prevent fatigue. After the online task was completed, subjects were debriefed about the study purposes. The whole session took about 40 minutes on average for each subject. Drinks and snacks were provided after the study.

\subsection{Results}

\subsubsection{Validity test}

Task completion time, responses to the task and mental effort were recorded. Based on the recorded data, visualization efficiency for each drawing was computed using a formula suggested by Huang et al.~\cite{huang09} to have an overall evaluation of task performance. Overall quality of each drawing was also computed using equation~\ref{eq3}. In this part of the study, dependent variables were time, accuracy, effort, efficiency and overall quality, while the independent variable was iteration number (of the force-directed algorithm).

\begin{table}[!t]
%% increase table row spacing, adjust to taste
\renewcommand{\arraystretch}{1.3}

\caption{Mean Values of Dependent Variables}
\label{tbl:average}
\centering
%% Some packages, such as MDW tools, offer better commands for making tables
%% than the plain LaTeX2e tabular which is used here.
\begin{tabular}{l|c|c|c|c}

\hline
Variable & C3  & C6 & C9 & C12\\
\hline
\hline
 Time (sec.) &  9.91   & 9.51   & 7.19   & 7.11 \\
 Effort      & 3.60    & 3.26   & 3.27   & 3.09 \\
 Accuracy    & 0.69    & 0.75   & 0.76   & 0.76 \\
 Efficiency  & -0.74   &-0.22   & 0.28   & 0.48 \\
\hline
 Overall quality  & -2.14   &0.04   & 1.02   & 1.08 \\
\hline
\end{tabular}
\end{table}

We obtained 10 (drawings) $\times$ 4 (conditions) $\times$ 35 (subjects) = 1440 experimental data entries for each dependent variable. The mean values were computed  and the results are shown in Table~\ref{tbl:average}. First, on performance measures, it can be seen that the subjects generally spent less time, exerted less effort and were more accurate when iteration number increased from condition c3, c6, c9 to c12. The performance efficiency also showed a clear increase across the conditions. In other words, the means of the performance measures were in good agreement with the pre-known overall quality.

 \begin{table*}[!t]
%% increase table row spacing, adjust to taste
\renewcommand{\arraystretch}{1.3}

\caption{Results of ANOVA with Post-Hoc Comparisons  }
\label{tbl:anova}
\centering
%% Some packages, such as MDW tools, offer better commands for making tables
%% than the plain LaTeX2e tabular which is used here.
\begin{tabular}{l|l|l|l}

\hline
Variable & \emph{F}(3,57)  & \emph{p} & Condition Pairs Detected Different (out of 6)\\
\hline
\hline
 Time        &  2.804  &0.048        &(c3, c9), (c3, c12)   \\
 Effort      &   7.491 &0.000     & (c3, c6), (c3, c9), (c3, c12)\\
 Accuracy    &   2.118 &0.108        & none \\
 Efficiency  &   7.442 &0.000     & (c3, c6), (c3, c9), (c3, c12)  \\
\hline
 Overall quality  &  28.596 &0.000& all pairs but (c9, c12) \\
\hline
%\multicolumn{4}{r}{\footnotesize{\emph{Notes}: \;\; *: $p < .05$; \;\; ***: $p < .001$}}\\
\end{tabular}
\end{table*}

To test whether these trends of changes were significant at the level of 0.05, we ran repeated ANOVA tests with post-hoc comparisons of the Least Square Difference (LSD) method on each of the dependent variables. The results are shown in Table~\ref{tbl:anova}. There was a significant main effect on time, effort and efficiency ($p < 0.05$), but not on accuracy ($p > 0.05$). Post-hoc comparisons revealed that these dependent variables had different levels of capacity in detecting condition differences. More specifically, out of the six condition pairs in total, time data only found that two pairs of the conditions were different; effort found three; and efficiency found three, while no difference was shown in accuracy between any pair of the conditions.
 
Second, on overall quality, the mean overall quality value was -2.14 for c3, 0.04 for c6, 1.02 for c9 and 1.08 for c12 (see the bottom row of Table~\ref{tbl:average}). These values showed that the measured overall quality increased while iteration number increased across the conditions from c3 to c12. To see whether this trend of increase was significant, we ran a repeated ANOVA with post-hoc comparisons. The results are shown in the bottom row of Table~\ref{tbl:anova}. The repeated ANOVA indicated that the main effect on overall quality was statistically significant, $F(3,57) = 28.596$, $p < 0.001$. Post-hoc comparisons indicated that all conditions were statistically different from each other, except the condition pair of c9 and c12.

\subsubsection{Predictability test}

Thirty-five subjects each viewed $30$ drawings for the shortest path task. The time they spent and their responses to the task and mental effort were recorded. In this part of the study, dependent variables were time, accuracy, effort and efficiency, while the predictor variable was overall quality. The obtained raw data were processed accordingly for each of the dependent variables. And the overall quality of each drawing was computed using equation~\ref{eq3}. In the end, we had $30$ data entries for each variable for data analysis.

 \begin{figure*}[t]
\centering
\includegraphics[width=4in]{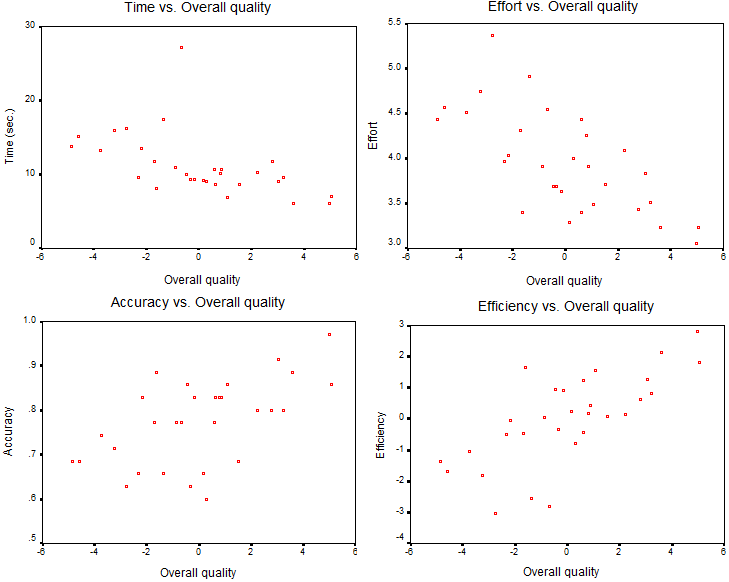}
\caption{Scatter diagrams between dependent variables and overall quality} 
\label{fig:pred}
\end{figure*}

We expected that the measured overall quality was negatively correlated with time and effort, and positively correlated with accuracy and efficiency. To test our hypotheses, we first plotted the scores of overall quality and the scores of each dependent variable as Cartesian coordinates to generate a scatter diagram, with overall quality on the horizontal axis and the dependent variable on the vertical axis. This was to have a general idea about the relationships between the two variables. The obtained diagrams are shown in Figure~\ref{fig:pred}. In these scatter diagrams, it appeared, as expected, that overall quality had a negative relationship with time and effort, and a positive relationship with accuracy and efficiency.

Then, we ran simple linear regression tests to see whether the observed relationships were statistically significant. We regressed each dependent variable on overall quality, and the results are shown in Table~\ref{pred}. In this table, $\beta$ is a standardized coefficient. The absolute value of it gives the size of effect that the predictor had on a dependent variable, while the sign implies the direction of that effect. According to common rules of thumb, effect size is small if $\beta$ is less than 0.10, is large if $\beta$ is more than 0.50, and is medium if $\beta$ is between 0.10 and 0.50.

\begin{table*}[!t]

\renewcommand{\arraystretch}{1.3}
\caption{Results of Simple Linear Regression Tests}
\label{pred}
\centering
\begin{tabular}{l|c|c|l|c}

\hline
Dependent Var. & Predictor  & $\beta$ & \emph{F}(1,28) & \emph{p}\\
\hline
\hline
Time & overall quality & -0.539 & 11.478 & 0.002\\
Effort & overall quality & -0.692 & 25.796 & 0.000\\
Accuracy & overall quality & 0.575 & 13.835 & 0.001\\
Efficiency & overall quality & 0.717 & 29.625  & 0.000\\
\hline
%\multicolumn{5}{r}{\footnotesize{\emph{Notes}:  \;\; **: $p < .01$; \;\; ***: $p < .001$}}\\
\end{tabular}
\end{table*}

The overall regression test of time was significant, $F(1,28) = 11.478$, $p < 0.01$. Time was negatively correlated with overall quality, $\beta = -0.539$. The overall regression test of effort was significant, $F(1,28) = 25.796$, $p < 0.001$. Effort was negatively correlated with overall quality, $\beta = -0.692$. The overall regression test of accuracy was significant, $F(1,28) = 13.835$, $p = 0.001$. Accuracy was positively correlated with overall quality, $\beta = 0.575$. The overall regression test of efficiency was significant, $F(1,28) = 29.625$, $p < 0.001$. Efficiency was positively correlated with overall quality, $\beta = 0.717$.

\subsection{Discussion}

Our data analysis for validity revealed that there was a significant overall difference shown in the data of performance measures including time, effort and efficiency, but not in the accuracy data. This indicated that the subjects had followed the instructions closely and did not compromise accuracy for speed in performing their tasks. The analysis also showed that the proposed overall quality measure was able to detect the quality difference between the drawings in the four conditions as expected. 

The further post-hoc pairwise comparisons revealed that the proposed measure was able to find more condition pairs being  different than any of the task performance and visualization efficiency measures did (see Table~\ref{tbl:anova}). This, on the one hand, indicated that the actual differences between conditions were small, which is good for the validity purpose. On the other hand, it indicated that the proposed overall quality measure was more sensitive to quality changes than performance measures. This should not be surprising if we consider that the proposed measure measures overall quality directly, while performance measures measure indirectly. In addition, performance measures require conduction of user studies, and factors associated with a user study could negatively affect performance measures in evaluating overall quality. More specifically, among other factors, human factors, methodological issues, data analysis methods and the choice of tasks can each have a certain role in the evaluation process, affecting the measurement in one way or another.

Our data analysis for predictability revealed that each of the dependent variables had a significant correlation with the predictor variable, overall quality. And the significant correlations came with large effect sizes as shown in the $\beta$ values. 

In summary, our tests demonstrated that given a graph, the proposed measure was able to not only differentiate drawings based on overall quality, but also significantly predict the performance of human graph comprehension. In other words, the proposed measure is a valid measure of overall quality. 

\section{General discussion}

In this paper, we proposed a measure that measures overall quality by combining individual aesthetic criteria into a single value. We have presented a user study that provides empirical evidence validating the proposed measure and verifying the predictive capacity of it. It is also shown in this particular study that the proposed measure performs better than performance measures in evaluating overall quality. It should be noted that this comparison was only to demonstrate the sensitivity of our new measure. And this finding should not be interpreted as an argument of one measure being used against another since they are essentially different kinds of measures serving for different purposes.

The proposed measure uses the $z$ scores of its aesthetic components to compute overall quality. Z scores have also been used for the purpose of measurement in the field of usability engineering~\cite{tullis08}. Usability is often considered as a multi-dimensional construct of user experience, in which each dimension is measured by a different metric. For example, ease of learning, ease of use, user preference and task satisfaction. The $z$ scores of these metrics are then combined to reflect the overall usability. Despite this, it is important to note that there is a fundamental difference between the overall usability and our overall layout quality. The former is empirical and is measured based on human user experience, while the latter is computational and is measured based on layout aesthetics.

One limitation of the proposed measure is that it assumes a linear relationship between overall quality and its associated component aesthetics, as implied in equation~\ref{eq3}.  This is clearly \newpage \noindent an  oversimplification and the reality can be more complex. For example, a study of Huang et al.~\cite{huang08} reveals that there exists a significant quadratic relationship between $crossRes$ and drawing quality measured by task response time. Although more studies are needed so that the actual relationships between them can be fully understood, the empirical evidence presented in this paper shows that our simplified version does give valid and useful insight into the relative overall quality between drawings.

Finally, our experiment has limitations. For example, we used only the shortest path search task. The obtained findings could be more widely applicable if different types of tasks were used~\cite{lee06}. Therefore, more studies are needed to verify the validity of our proposed measure.

 %Must end the environment

\end{document}